\documentclass[12pt]{article}
\usepackage{epsf}

\begin{document}

\newcommand{\gtrsim}{\mathop{}_{\textstyle \sim}^{\textstyle >}}
\newcommand{\lesssim}{\mathop{}_{\textstyle \sim}^{\textstyle <} }

\newcommand{\rem}[1]{{\bf #1}}

\renewcommand{\thefootnote}{\fnsymbol{footnote}}
\setcounter{footnote}{0}
\begin{titlepage}

\def\thefootnote{\fnsymbol{footnote}}

\begin{center}

\hfill TU-785\\
\hfill March, 2007\\

\vskip .75in

{\Large \bf 
Big-Bang Nucleosynthesis with Long-Lived Charged Slepton
}

\vskip .75in

{\large
$^{(a)}$Masahiro Kawasaki, $^{(b)}$Kazunori Kohri, and $^{(c)}$Takeo Moroi
}

\vskip 0.25in

{\em $^{(a)}$Institute for Cosmic Ray Research,
University of Tokyo, Kashiwa 277-8582, JAPAN}

\vskip 0.2in

{\em $^{(b)}$Physics Department, Lancaster University, Lancaster LA1 4YB,
UK}

\vskip 0.2in

{\em $^{(c)}$Department of Physics, Tohoku University,
Sendai 980-8578, JAPAN}

\end{center}
\vskip .5in

\begin{abstract}

We consider constraints on long-lived charged scalar leptons
$\tilde{l}^\pm$ in supersymmetric models, where gravitino is the
lightest superparticle.  We study the decay and hadronization
processes of $\tilde{l}^\pm$.  We also take into account the
significant enhancement of ${\rm ^6Li}$ production due to the
formation of the bound-state $({\rm ^4He}\tilde{l}^-)$; for this
purpose, we use the reaction rate given by the most recent calculation
based on coupled-channel method.

\end{abstract}

\end{titlepage}

\renewcommand{\thepage}{\arabic{page}}
\setcounter{page}{1}
\renewcommand{\thefootnote}{\#\arabic{footnote}}
\setcounter{footnote}{0}

Cosmology provides important constraints on physics beyond 
the standard model.  One of the important examples is constraints on 
long-lived particles from big-bang nucleosynthesis (BBN); since 
the theoretical predictions of the standard BBN scenario are 
in reasonable agreements with
observation of light-element abundances, we obtain constraints on
long-lived particles whose decay may induce photo- and
hadro-dissociation processes.

Here, as a physics beyond the standard model, we consider low energy
supersymmetry (SUSY), which is strongly motivated as a prominent
solution to serious problems in the standard model, like the hierarchy
problem, the dark-matter problem, and so on.  In the framework of SUSY
models, we assume that one of the charged sleptons is the lightest
particle among the superpaticle in the
minimal-supersymmetric-standard-model (MSSM) sector, and that the
gravitino is the lightest superparticle (LSP).  Such a mass spectrum is
easily realized; probably the most famous example is gauge-mediated SUSY
breaking models \cite{GMSB}.  Thus, such a scenario has been attracted
much attention, in particular, in the context of gravitino LSP
\cite{Moroi:1993mb,Feng:2003xh,Feng:2003uy,Ellis:2003dn,
Feng:2004zu,Feng:2004mt,Roszkowski:2004jd,Cerdeno:2005eu}.  Then, the
lightest slepton $\tilde{l}$ is usually long-lived and the decay of
$\tilde{l}$ may spoil the success of BBN.  Indeed, the BBN scenario with
long-lived charged sleptons has been discussed in several works
\cite{Feng:2004zu,Feng:2004mt,Steffen:2006hw}.

In this letter, we reconsider effects of long-lived charged slepton on
BBN.  We perform a detailed study on the decay processes of slepton
and the hadronization processes of the decay products.  In addition,
we also take into account the effect of bound-state formation;
recently, it has been pointed out that long-lived charged slepton may
form bound states with light elements
\cite{Pospelov:2006sc,Kohri:2006cn,BSeffects}, which may significantly
change the BBN reaction rates~\cite{Pospelov:2006sc}.  In particular
we take a serious attitude towards the bound-state effect by (${\rm
^4He} \tilde{l}^-$) pointed out by \cite{Pospelov:2006sc} (which is
called $\tilde{l}^-$-catalyzed process) and include it into our
analysis, taking into account the result of the most recent precise
calculation of the bound-state effect \cite{Hamaguchi:2007mp}.

Let us first summarize the basic properties of the lightest slepton
$\tilde{l}$ and the gravitino (which is denoted as $\psi_\mu$
hereafter).  Since the gravitino is the only superparticle lighter
than $\tilde{l}$, the dominant decay mode of $\tilde{l}$ is
$\tilde{l}\rightarrow l\psi_\mu$.  The decay rate for this process is
given by
\begin{eqnarray}
 \Gamma_{\tilde{l}} = \frac{m_{\tilde{l}}^5}{48\pi m_{3/2}^2 M_*^2}
  \left( 1 - \frac{m_{3/2}^2}{m_{\tilde{l}}^2} \right)^4,
  \label{Gamma(2body)}
\end{eqnarray}
where $m_{\tilde{l}}$ and $m_{3/2}$ are slepton mass and gravitino mass,
respectively, and $M_*\simeq 2.4\times 10^{18}\ {\rm GeV}$ is the
reduced Planck scale.  When the gravitino is much lighter 
than $\tilde{l}$, we obtain
\begin{eqnarray}
  \tau_{\tilde{l}} \simeq 5.7\times 10^4 \ {\rm sec} \times
  \left( \frac{m_{\tilde{l}}}{100\ {\rm GeV}} \right)^{-5}
  \left( \frac{m_{3/2}}{1\ {\rm GeV}} \right)^{2}.
\end{eqnarray}
For the case where the next-to-the-lightest superparticle (NLSP) is
selectron $\tilde{e}$, the daughter particles are stable.  For other
cases, on the contrary, the final-state lepton also decays; such decay
process is sometimes important.  In particular, when stau is the NLSP,
$\tau$-lepton in the final state decays subsequently and produces
pion(s).  As we will discuss later, such pions become a source of
$p\leftrightarrow n$ conversion.  In addition, when $\mu$- and
$\tau$-lepton decays, neutrinos carry away some amount of energy, which
affects the total amount of visible energy emitted by the decay of
$\tilde{l}$.

In our analysis, we study the decay processes of unstable leptons (in
particular, that of $\tau$-lepton) using PYTHIA package
\cite{Sjostrand:2006za} and derive energy distributions of the decay
products.  For the decay of $\tau$-lepton, for example, the averaged
number of charged pions produced by the decay of $\tau$-lepton is
0.94, which changes the ${\rm ^4He}$ abundance as we will see later.
With the result given by PYTHIA analysis, we also calculate the
(averaged) visible energy emitted from $\tilde{l}$ taking into account
the energy loss due to the neutrino emission.

Charged sleptons may also decay into 3- and/or 4-body final states as
$\tilde{l}\rightarrow l\psi_\mu Z^{(*)}$, followed by the decay of
$Z^{(*)}$.\footnote
{Here, we assume that the sleptons are right-handed.  In such a case,
3- and 4-body decay processes via the emission of $W^\pm$ are
irrelevant.}
Although 3- and 4-body decay processes are sub-dominant, they are
important for the study of hadro-dissociation processes.  When $Z^{(*)}$
decays into $q\bar{q}$ pair, protons, neutrons, and pions are emitted
after hadronizing final-state quarks.  These hadrons induce
hadro-dissociation and $p\leftrightarrow n$ conversion processes, which
affect abundances of light elements.  (For details, see, for example,
\cite{KKM04}.)  In order to study the effects of hadro-dissociation and
$p\leftrightarrow n$ conversion processes, it is necessary to know the
hadronic branching ratio of $\tilde{l}$ as well as the spectra of
final-state hadrons.  We have also calculated the decay rate for the
processes $\tilde{l}\rightarrow l\psi_\mu q\bar{q}$.  Detailed
information about the 3- and 4-body decay processes can be found in
\cite{Steffen:2006hw}; we found that our results are consistent with
those shown in \cite{Steffen:2006hw}.  We have also calculated the
invariant-mass and energy distributions of final-state quark pair.
Then, using PYTHIA package, we calculate energy distributions of $p$,
$n$, and $\pi^\pm$.

Once the decay rate, total visible energy emitted by the decay of
$\tilde{l}$, and spectra of $p$, $n$ and $\pi^\pm$ are obtained, we
calculate the light-element abundances taking account of photo- and
hadro-dissociation processes as well as $p\leftrightarrow n$
conversion processes;\footnote
{We have omitted contributions from Kaons as discussed in
\cite{KKM04}.} 
we follow the procedure given in \cite{KKM04} in order to calculate
the abundances of ${\rm D}$, ${\rm ^3He}$, ${\rm ^4He}$, ${\rm ^6Li}$
and ${\rm ^7Li}$ for processes except for $({\rm
^4He}\tilde{l}^-)+{\rm D}\rightarrow {\rm ^6Li}+\tilde{l}^-$.

One of the important new ingredients in the current study is the
$\tilde{l}^-$-catalyzed process for ${\rm ^6Li}$ production
\cite{Pospelov:2006sc}.\footnote
{See also \cite{Kohri:2006cn,BSeffects} for recent progress in this
field.}
In the standard BBN scenario, the production rate of ${\rm ^6Li}$ is
very tiny because the E1 transition process for ${\rm D}({\rm
^4He},\gamma){\rm ^6Li}$ process is extremely suppressed, resulting in
very small primordial abundance of ${\rm ^6Li}$ which is consistent
with observation.  Recently, however, it was claimed that the ${\rm
^6Li}$ abundance may become much larger than the standard BBN
prediction if there exists a long-lived heavy charged particle
\cite{Pospelov:2006sc}.  This is due to the enhancedment of the
process
\begin{eqnarray}
  ({\rm ^4He}\tilde{l}^-)+{\rm D}\rightarrow {\rm ^6Li}+\tilde{l}^-,
  \label{Li6_BS}
\end{eqnarray}
where $({\rm ^4He}\tilde{l}^-)$ denotes the bound state formed by
${\rm ^4He}$ and $\tilde{l}^-$.  Importantly, for this process,
accurate calculation for the reaction rate is performed with the use
of coupled-channel method which is known to accurately describe other
three-body systems \cite{Hamaguchi:2007mp}.  Thus, we adopt the
results of \cite{Hamaguchi:2007mp} and take into account the
bound-state effect in the calculation of ${\rm ^6Li}$ abundance.

In order to estimate the number density of $({\rm ^4He}\tilde{l}^-)$
bound state $n_{\rm BS}$, we solve the Saha-type equation
\cite{Kolb:1990vq}:\footnote
{For more detailed discussion, see~\cite{Kohri:2006cn,KT07}.}
\begin{eqnarray}
  n_{\rm BS} = 
  \left( \frac{m_{\rm ^4He} T}{2\pi} \right)^{-3/2} e^{E_{\rm B}/T}
  ( n_{\rm ^4He} - n_{\rm BS} )
  ( n_{\tilde{l}^-} - n_{\rm BS} ),
  \label{SahaEq}
\end{eqnarray}
where $E_{\rm B}\simeq 337.33\ {\rm keV}$ \cite{Hamaguchi:2007mp} is the
binding energy, and $n_{\rm ^4He}$ and $n_{\tilde{l}^-}$ are total
number densities of ${\rm ^4He}$ and $\tilde{l}^-$ (which include those
in the bound state), respectively.  (In this letter, $n_X$ is used for
the number density of particle $X$.)  Here, we assume that the process
${\rm ^4He}+\tilde{l}^-\leftrightarrow({\rm ^4He}\tilde{l}^-)$ is in
chemical equilibrium.  Then, the total reaction rate $R$ for ${\rm
^6Li}$ production by D-$^4$He collision is calculated as
\begin{eqnarray}
 R = \frac{n_{\rm BS}}{n_{\rm ^4He}} R_{\rm BS} + 
  \left( 1 - \frac{n_{\rm BS}}{n_{\rm ^4He}} \right) R_{\rm SBBN},
\end{eqnarray}
where $R_{\rm BS}$ is given by Eq.\ (4.3) in \cite{Hamaguchi:2007mp} and
$R_{\rm SBBN}$ is the standard-BBN value.

Once the light-element abundances are theoretically calculated, we
compare them with observational constraints and derived 95\ \%
C.L. constraints.  As observational constraints on the primordial
abundances of light elements, we adopt those used in
\cite{Kanzaki:2006hm} except for $Y_{\rm p}$ and $(n_{\rm ^3He}/n_{\rm
D})_{\rm p}$:
\begin{eqnarray}
  (n_{\rm D}/n_{\rm H})_{\rm p} &=& (2.82\pm 0.26) \times 10^{-5},
  \\
   (n_{\rm ^3He}/n_{\rm D})_{\rm p} &<& 0.83+0.27,
  \\
   Y_{\rm p} &=& 0.2516 \pm 0.0040,
  \\
  \log_{10}(n_{\rm ^7Li}/n_{\rm H})_{\rm p} &=& -9.63 \pm 0.06 \pm 0.3.
  \\
  (n_{\rm ^6Li}/n_{\rm ^7Li})_{\rm p} &<& 0.046\pm 0.022 + 0.084,
  \label{Li6/Li7}
\end{eqnarray}
where the subscript ``p'' is for primordial value (just after BBN),
and $Y_{\rm p}$ is the primordial mass fraction of ${\rm ^4He}$. For
the center value of $Y_{\rm p}$, we have adopted a quite recent value
reported in \cite{Izotov:2007ed} in which the authors used new data of
HeI emissivites,\footnote
{See also recent value of $Y_{\rm p}$ reported in
\cite{Peimbert:2007vm} where the authors adopted larger errors
(0.0028) than that of \cite{Izotov:2007ed}.}
and conservatively added larger errors of $0.0040$ as shown in
\cite{Fukugita:2006xy}.  For $(n_{\rm ^3He}/n_{\rm D})_{\rm p}$ we
have used most newly-reported values of D and $^{3}$He abundances
observed in the protosolar cloud \cite{GG03}.

Now, we are at the position to show our results.  First, we consider
the case where there is no entropy production after the freeze out of
$\tilde{l}$.  In this case, the abundance of $\tilde{l}$ is thermally
determined; then, for right-handed slepton, the yield variable (which
is defined as $Y_{\tilde{l}} \equiv [n_{\tilde{l}}/s]_{t\ll
\tau_{\tilde{l}}}$, with $s$ being entropy density) is estimated
as \cite{Fujii:2003nr}
\begin{eqnarray}
  Y_{\tilde{l}}
  \simeq 7 \times 10^{-14} \times
  \left( \frac{m_{\tilde{l}}}{100\ {\rm GeV}} \right).
  \label{Y_sl(thermal)}
\end{eqnarray}
Notice that, in the solving Eq.\ (\ref{SahaEq}), $n_{\tilde{l}^-}$ is
given by
\begin{eqnarray}
 n_{\tilde{l}^-}=\frac{1}{2}sY_{\tilde{l}}e^{-t/\tau_{\tilde{l}}}.
\end{eqnarray}
since $n_{\tilde{l}^-} = n_{\tilde{l}^+} = 0.5 n_{\tilde{l}}$.

\begin{figure}[t]
    \centerline{\epsfxsize=0.6\textwidth\epsfbox{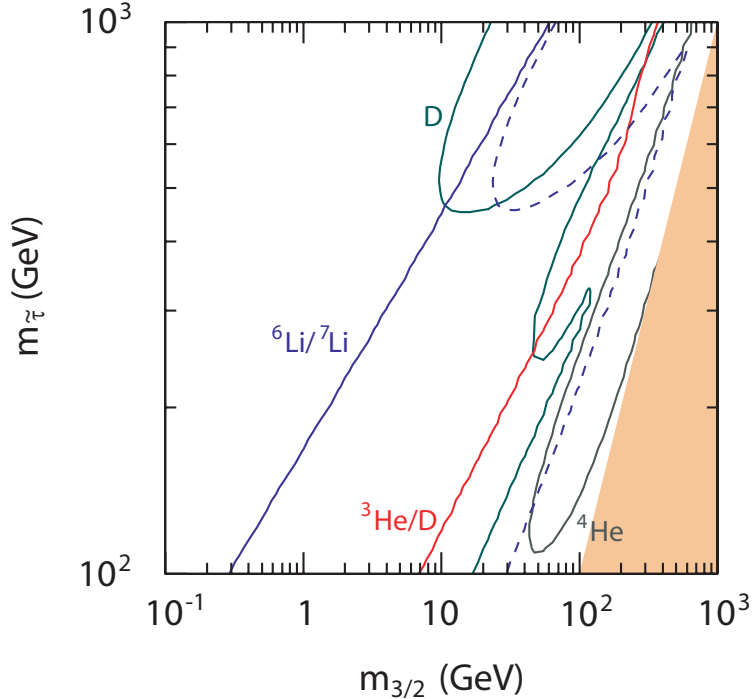}}
    \caption{Constraints on $m_{3/2}$ vs.\ $m_{\tilde{\tau}_R}$ plane.
      The dashed line indicates the constraint from 
      $(n_{\rm ^6Li}/n_{\rm ^7Li})_{\rm p}$ without taking account of the
      $\tilde{l}^-$-catalyzed process.  We have shaded the region where
      $\tilde{\tau}_R$ becomes lighter than gravitino.}
  \label{fig:Yth}
\end{figure}

With the thermal abundance, we perform our analysis for the case where
$\tilde{l}$ is $\tilde{\tau}_R$.  The result is shown in Fig.\
\ref{fig:Yth}.  For comparison, we also show the result without
including the $\tilde{l}^-$-catalyzed process (\ref{Li6_BS}).  As
expected, constraints from light elements other than ${\rm ^6Li}$ is
unaffected by the inclusion of the $\tilde{l}^-$-catalyzed process
(\ref{Li6_BS}); those constraints are more or less consistent with the
results obtained in earlier analysis
\cite{Feng:2004zu,Feng:2004mt,Steffen:2006hw}.  As one can see,
however, significant amount of the parameter region is excluded with
the overproduction of ${\rm ^6Li}$ once the $\tilde{l}^-$-catalyzed
process is considered.  For the case where $m_{\tilde{l}}=100\ {\rm
GeV}$, for example, the upper bound on the gravitino mass becomes
$\sim 0.2\ {\rm GeV}$, which was $\sim 7\ {\rm GeV}$ before the
process (\ref{Li6_BS}) is taken into account.

\begin{figure}[t]
  \centerline{\epsfxsize=0.6\textwidth\epsfbox{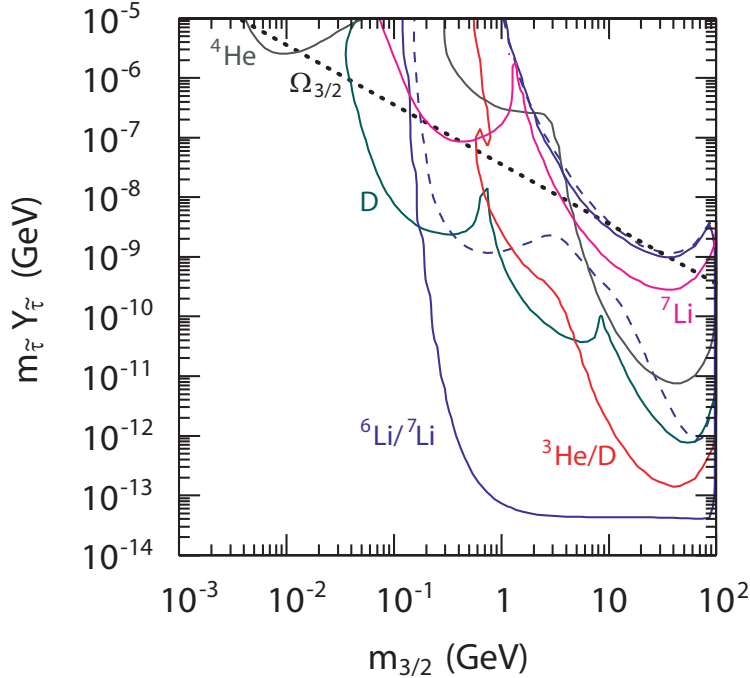}}
  \caption{Upper bounds on $m_{\tilde{\tau}_R}Y_{\tilde{\tau}_R}$ as
    functions of $m_{3/2}$ for $m_{\tilde{\tau}_R}=100\ {\rm GeV}$.
    The solid lines are bounds from various light-element abundances,
    while the dashed line is the upper bound from $(n_{\rm
    ^6Li}/n_{\rm ^7Li})_{\rm p}$ without taking account of the
    $\tilde{l}^-$-catalyzed process.  Thick dotted line shows the contour
    where the density parameter of the gravitino produced by the decay of
    stau becomes as large as that of cold dark matter.}
  \label{fig:M100}
\end{figure}

\begin{figure}[t]
  \centerline{\epsfxsize=0.6\textwidth\epsfbox{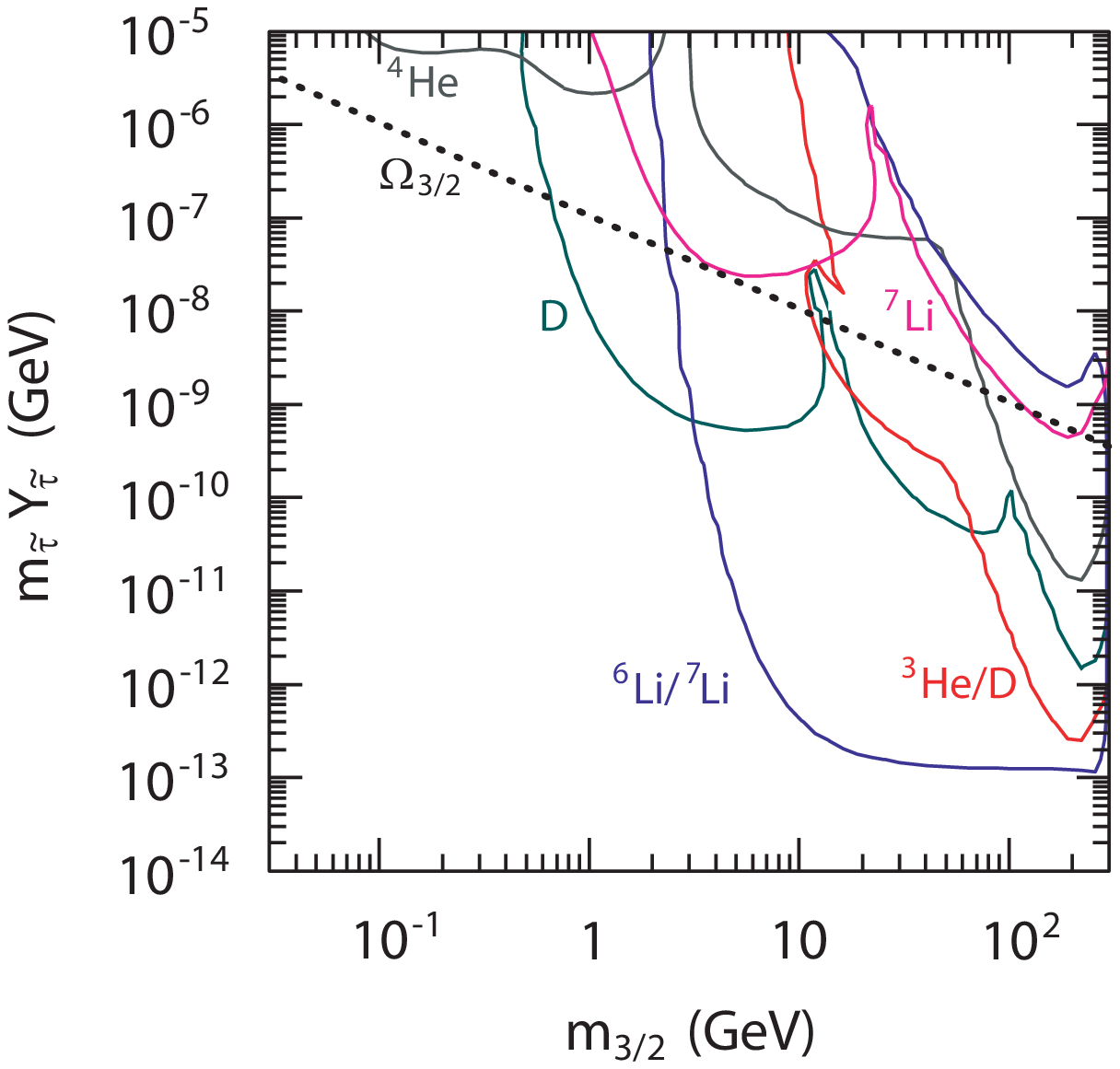}}
  \caption{Same as Fig.\ \ref{fig:M100}, except for 
    $m_{\tilde{\tau}_R}=300\ {\rm GeV}$.}
  \label{fig:M300}
\end{figure}

So far, we have assumed that the primordial abundance of $\tilde{l}$
is thermally determined.  With non-standard evolution of the universe,
however, $Y_{\tilde{l}}$ may deviate from the thermal value given in
Eq.\ (\ref{Y_sl(thermal)}).  Thus, we also treat the primordial
abundance of $\tilde{l}$ as a free parameter and derive upper bound on
$Y_{\tilde{l}}$.  For $\tilde{\tau}_R$-NLSP, the result is shown in
Fig.\ \ref{fig:M100}, taking $m_{\tilde{\tau}_R}=100\ {\rm GeV}$.  As
one can see, the $\tilde{l}^-$-catalyzed process significantly changes
the upper bound; for $m_{\tilde{\tau}_R}=100\ {\rm GeV}$, for example,
the very stringent constraint is obtained when the gravitino mass is
larger than $O(0.1\ {\rm GeV})$, which corresponds to
$\tau_{\tilde{l}}\gtrsim O(10^3\ {\rm sec})$.  We have also considered
the case with $m_{\tilde{\tau}_R}=300\ {\rm GeV}$, and the result is
shown in Fig.\ \ref{fig:M300}.  For $m_{\tilde{\tau}_R}=300\ {\rm
GeV}$, the constraints from the hadro-dissociation processes are more
stringent than those for $m_{\tilde{\tau}_R}=100\ {\rm GeV}$.  This is
mainly because, for the case of $m_{\tilde{\tau}_R}=300\ {\rm GeV}$,
the decay mode $\tilde{\tau}_R\rightarrow \tau Z$ becomes more
effective, resulting in larger number of $p$ and $n$.

We have also studied another cases that the NLSP is the right-handed
selectron $\tilde{e}_R$.  In this case, the constraints slightly
change.  First, as we mentioned earlier, in the dominant decay mode,
$\tilde{e}_R$ decays into gravitino and electron, which are both
stable.  Thus, contrary to the $\tilde{\tau}_R$-case, the pion 
production is only via three- and/or four-body decay processes, 
and hence is significantly suppressed.  
Consequently, the constraint from the
$p\leftrightarrow n$ conversion is weakened.  Second, the visible
energy emitted by the slepton decay is larger for the selectron-NLSP
case since there is no energy loss due to the neutrino emission (in
the two-body decay process).  As a result, the constraint from the
photo-dissociation process becomes more stringent than the stau-NLSP
case.  We checked that the upper bound on $Y_{\tilde{l}}$ from the
ratio $(n_{\rm ^3He}/n_{\rm D})_{\rm p}$ becomes smaller by factor
$\sim 3$ compared to the stau-NLSP case.

Finally, we comment on the bound-state effects on other processes.  In
our analysis, we consider only the $\tilde{l}^-$-catalyzed process for
the ${\rm ^6Li}$ production.  In fact, reaction rates of other processes
for which the E1 transition is effective (like ${\rm T}({\rm
^4He},\gamma){\rm ^7Li}$ and ${\rm ^3He}({\rm ^4He},\gamma){\rm ^7Be}$)
may be also enhanced by the bound-state formation.  Unfortunately, for
those processes, we could not find reliable calculation of the reaction
rates.  However, we expect that the constraint from ${\rm ^7Li}$ is less
stringent than that from ${\rm ^6Li}$ even if the bound-state effect is
taken into account for ${\rm ^7Li}$ and ${\rm ^7Be}$ productions; there
are several reasons why we consider so.  First, observationally, ${\rm
^6Li}$ is rarer than ${\rm ^7Li}$, as one can see in Eq.\
(\ref{Li6/Li7}).  Thus, the constraint on the model is usually much more
sensitive to the change of ${\rm ^6Li}$ rather than that of ${\rm
^7Li}$.  In addition, reaction rates for the processes ${\rm T}(({\rm
^4He}l^-),l^-){\rm ^7Li}$ and ${\rm ^3He}(({\rm ^4He}l^-),l^-){\rm
^7Be}$ should acquire extra suppressions compared to the process
(\ref{Li6_BS}); for the formar, one should keep in mind that the number
density of ${\rm T}$ during BBN is about two orders of mangnitude
smaller than that of ${\rm D}$ while, for the latter, the Coulomb
suppression becomes much more significant compared to the process
(\ref{Li6_BS}).  Furthermore, we have also performed numerical
calculation by using the enhancement factor presented in
\cite{Pospelov:2006sc} for ${\rm ^7Li}$ and ${\rm ^7Be}$ productions.
(Notice that, for ${\rm ^6Li}$ production process (\ref{Li6_BS}), the
formula given in \cite{Pospelov:2006sc} predicts reaction rate which is
one order of magnitude larger than the result of the detailed
calculation given in \cite{Hamaguchi:2007mp}.)  Then we have checked
that, even with these modifications of reaction rates, constraint from
${\rm ^7Li}$ is much less stringent than that from ${\rm ^6Li}$.  Thus,
we expect that the bound-state effects on the ${\rm ^7Li}$ and ${\rm
^7Be}$ production processes do not change our main conclusion.

{\sl Note Added:} After the completion of this work, we found the paper
\cite{Bird:2007ge} which discusses possible effects of the formation of
bound states $({\rm ^7Li}\tilde{l}^-)$ and $({\rm ^7Be}\tilde{l}^-)$.
Such bound states, however, do not have significant effect on ${\rm
^6Li}$ formation and hence our main conclusion is unchanged.

{\sl Acknowledgements:} This work was supported in part by PPARC grant,
PP/D000394/1, EU grant MRTN-CT-2006-035863, the European Union through
the Marie Curie Research and Training Network "UniverseNet"
(MRTN-CT-2006-035863) (KK).

\end{document}